\documentstyle[pra,preprint,aps]{revtex}
\renewcommand{\baselinestretch}{1}
\begin{document}
\title{Remote state preparation and measurement of single photon
 }
\author{Arun K. Pati$^{(1)}$}
\address{Institute of Physics, Sainik School Post, Bhubaneswar-751005, India} 
\address{$^{(1)}$School of Informatics, University of Wales, Bangor LL 57 
1UT, UK}

\maketitle
\def\ra{\rangle}
\def\la{\langle}
\def\ver{\arrowvert}
\begin{abstract}
Quantum information theory has revolutionized the way in which 
information is processed using quantum resources such as 
entangled states, local operations  
and classical communications. Two important protocols in quantum 
communications are quantum 
teleportation and remote state preparation. In quantum teleportation 
neither the sender nor the receiver know the identity of a state. 
In remote state preparation the sender knows 
the state which is to be remotely prepared without ever physically 
sending the object or the complete classical description of it. 
Using one unit of entanglement and one classical bit Alice can 
remotely prepare a photon (from special ensemble) of her choice at 
Bob's laboratory. In remote
state measurement Alice asks Bob to simulate any single particle 
measurement statistics on an arbitrary photon. In this talk we will present
these ideas and discuss the latest developments and future open problems.
\end{abstract}

\vskip .5cm

PACS           NO:    03.67.-a, 03.65.Bz\\

email:akpati@iopb.res.in\\

\vskip 1cm


\par

\section{Introduction}
Quantum information theory is a vast area of interdisciplinary research 
which includes quantum computing, quantum complexity, quantum communication 
protocols, quantum cryptography, quantum entanglement, and so on \cite{nc}. 
In quantum communication protocols how to process the inaccessible quantum 
information contained in an unknown state using quantum entanglement 
and classical channels is an ongoing important area of research.
The most cited example is quantum 
{\em teleportation}  of an {\em unknown} state \cite{cb} from one place 
to another without ever physically sending the particle. Any quantum 
state can be teleported from Alice to Bob provided they share a maximally 
entangled pair and allow to communicate classically. Alice carries out 
a Bell-state measurement on the input state and one half of the entangled pair 
and sends the measurement outcomes via a classical channel to Bob. 
Then he performs a unitary operation on his particle to get the original 
state. In the process the original state
is destroyed and a replica appears in accordance with no-cloning principle
\cite{wz,dd}. In a similar manner, one may interpret that since the original
is being destroyed at sender's location it must appear somewhere else (i.e. at
receiver's location) in accordance with no-deletion principle \cite{pb}. Thus
quantum information is `robust' in some sense (and `fragile' too)!
If Alice and Bob do not share maximally entangled state rather non-maximally
entangled state, then via {\em non-maximally entangled} measurement 
and classical communication Alice can teleport a single photon in a 
probabilistic manner \cite{pankaj}. 

However, if Alice has the complete classical information about a state
she need not do teleportation. Instead, she can help Bob to prepare a 
quantum state at a remote location using prior entanglement and classical
communication. It was shown that indeed Alice can prepare special class of
qubits (either from polar or equatorial great circle) using one maximally
entangled pair and
one classical bit of communication \cite{akp1}. In section I, I discuss 
mainly the exact remote state preparation of single photon. 
In section III, I discuss remote state measurement (RSM) of arbitrary 
photon via projection and generalised POVM measurements. In section IV, 
I present why Alice cannot ask Bob to simulate two particle measurement 
on an unknown photon with one unit of entanglement and one classical bit. 
In section V, I will mention recent developments in the context of
remote state preparation and its generalisation to higher dimensional
quantum system.

\section{Remote state preparation of a special photon}

In this section I discuss the remote state preparation protocol
of a single photon chosen from a particular ensemble. The particular
ensemble is known both to Alice and Bob, but the parameters describing
the state in the ensemble is {\em know} only to Alice but 
{\em unknown to Bob}. 
This protocol requires only one classical bit to be communicated from
Alice to Bob provided they share an Einstein-Podolsky-Rosen (EPR) pair. 
Here, we do not require a Bell-state measurement; only a local unitary
operation and single particle von Neumann measurement is necessary. 
Moreover we do not need the physical presence of a qubit at Alice's 
location. The possibility that a photon chosen from equatorial 
or polar great circles on a Poincare sphere can be remotely prepared 
arises from the isotropic nature of EPR state and impossibility 
of complementing an unknown state.
 
Consider a photon in a pure state $\ver\Psi \ra$.
An arbitrary photon can be represented as a linear superposition of two 
distinct polarisation states such as
\begin{equation}
\ver \Psi \ra = \alpha \ver H \ra + \beta \ver V \ra,
\end{equation}
where we can choose  $\alpha$ to be real and $\beta$ to be a complex number,
in general (up to $U(1)$ equivalence classes of states). $\ver H \ra$
and $\ver V \ra$ represents horizontal and vertical polarisation states 
of a photon. They are also called computational basis states, because they
can represent classical information such as `$0$' and `$1$'. The single 
photon state in (1) is a quantum bit (qubit) and 
can be represented by a point on the Poincare sphere $S^2$. Such a qubit
is realized only in quantum world but not in the classical world.
The qubit $\ver \Psi \ra$ is {\em known} to  Alice and  {\em unknown} to Bob. 

Imagine that Alice and Bob are far away from each other. Alice wants a 
photon in the state $\ver \Psi \ra$ at Bob's place. One way would be to
prepare a photon in her lab and send through an optical fiber
over to Bob's lab. Alternately, she can send the classical description of
the photon over an ordinary communication channel (telephone line) and Bob
prepares himself accordingly. In the first case, some one else on the way
can take the photon that Alice has sent and in the second case, to 
transmit two real numbers
one needs to send infinite number of bits, in principle. From communication
point of view this would be very expensive. However, both the problems can be
over come if Alice and Bob have shared previously one half of the photons 
from an EPR source, which is given by
\begin{equation}
\ver \Psi^- \ra_{AB} = {1 \over \sqrt 2}(\ver H \ra_A \ver V \ra_B -
\ver V \ra_A \ver H \ra_B).
\end{equation}
This type of entangled state of photons can be created by parametric
down-conversion process where an input photon of given frequency 
(called pump photon) can decay inside a non-linear crystal into 
two photons with maximal quantum correlation.  

Now the protocol goes like this. Alice is having first particle and Bob is 
having the second. Since Alice knows the state she can 
chose to measure her photon in any  basis she wants. Furthermore, 
she knows how to relate the computational basis sates $\{ \ver H \ra, 
\ver V \ra \}$ and the arbitrary ``qubit basis''$\{ \ver \Psi \ra, 
\ver \Psi_{\perp} \ra \}$ in a unitary manner. The unitary operator is 
the standard $SU(2)$ operator. 
This is given by
\begin{eqnarray}
\ver \Psi \ra=
U(\alpha, \beta) \ver H \ra = \alpha \ver H \ra  + \beta \ver V  \ra
 \nonumber\\
\ver \Psi_{\perp} \ra=
U(\alpha, \beta) \ver V \ra = \alpha \ver V \ra - 
\beta^* \ver H \ra.
\end{eqnarray}

An important property of the EPR state $\ver \Psi^- \ra_{AB}$ is that it is 
invariant under local unitary operator $U_A \otimes U_B$ operation where 
the same $U$ acts on both the subsystems, i.e.,
\begin{eqnarray}
U_A(\alpha, \beta) \otimes U_B(\alpha, \beta) \ver \Psi^- \ra_{AB} = 
{1 \over \sqrt 2}[ \ver \Psi \ra_A \ver \Psi_{\perp} \ra_B
 - \ver \Psi_{\perp} \ra_A \ver \Psi \ra_B ] =\ver \Psi^- \ra_{AB}.
\end{eqnarray}

Now another remarkable property follows from the above invariance nature of
EPR singlet is that {\em if a subsystem undergoes an evolution, then the other 
subsystem undergoes a de-evolution} or vice versa.   It is something which 
is really counter intuitive and has no classical analog! It is 
expressed by the following equation:
\begin{eqnarray}
U_A(\alpha, \beta) \otimes I_B \ver \Psi^- \ra_{AB} = 
I_A \otimes U_B(\alpha, \beta)^{\dagger}\ver \Psi^- \ra_{AB}.
\end{eqnarray}
This says that when both the photons are in an EPR state, then if one 
subsystem goes forward in time and the other one is silent (done nothing), 
it is equivalent to first subsystem being silent and the second goes 
back reverse in time. However, this evolution cannot be seen at 
individual level, because the state of either photon is completely 
unpolarized (i.e. a random density matrix $I/2$). The evolution leading to 
state preparation is possible only after local measurement and 
sending the classical information. To me this lies at the heart of 
remote state preparation of special class of photons. 

Now Alice applies $U_A(\alpha, \beta)^{\dagger}$ to her particle and as a 
result the state becomes
\begin{eqnarray}
U_A(\alpha, \beta)^{\dagger} \otimes I_B \ver \Psi^- \ra_{AB} = 
{1 \over \sqrt 2}[ \ver H \ra_A \ver \Psi_{\perp} \ra_B
 - \ver V \ra_A \ver \Psi \ra_B ] .
\end{eqnarray}
She performs a von Neumann projection onto horizontal and vertical basis 
and sends one classical bit to Bob. If her outcome is $\ver H \ra$ 
Bob's photon would be in $\ver \Psi_{\perp} \ra$ and if she gets $\ver V\ra$
then Bob's photon would be in the desired state $\ver \Psi \ra$.
For example, if Alice  chooses to prepare a photon from polar great 
circle, i.e., $\ver \Psi \ra = \alpha \ver H \ra + \beta \ver V \ra $ 
(with $\alpha$ and $\beta$ both are real), Bob will apply 
$ i \sigma_y$  to $\ver \Psi_{\perp} \ra= \beta \ver H \ra - 
\alpha \ver V \ra $
or do nothing after receiving the classical information from Alice. 
Alternatively, if Alice chose to prepare a photon from equatorial circle on 
Poincare sphere such as $\ver \Psi \ra = \frac{1}{\sqrt 2} 
( \ver H \ra + e^{i \phi} \ver V \ra )$ then in this case Bob can get 
$ \ver \Psi \ra $ from $\ver \Psi_{\perp} \ra= \frac{1}{\sqrt 2}
( \ver H \ra - e^{i \phi} \ver V \ra )$ by applying $\sigma_z$ or do
nothing after receiving one classical bit. Thus Alice can prepare  
any photon either from polar or equatorial circle using this protocol 
at a remote location.

But why cannot Alice prepare any arbitrary photon state even though she
has complete knowledge. Here comes Bob's knowledge in manipulating photon
states! Similar to no-cloning and
no-deleting principles, we know that an arbitrary unknown state cannot 
be complemented \cite{akp,bhw,gp}. Here complementing operation means 
creating a state that is orthogonal to the given state. Though we can 
design a device (called NOT gate) which can take 
$\ver H \ra \rightarrow \ver V \ra$ and $\ver V \ra \rightarrow \ver H \ra$ 
there is no universal NOT gate which can take an unknown qubit 
$\ver \Psi \ra \rightarrow \ver \Psi_{\perp} \ra$ as it involves an 
{\em anti-unitary} operation. Anti-unitary
operations correspond to improper rotations and cannot be implemented by
physical operations (called Completely Positive Maps).
Thus, Bob cannot convert a photon in an orthogonal state (which he gets 
half of the time) since it is {\em unknown} to him. 
Thus, a doubly infinity of bits of information (corresponding to two real
numbers \cite{rj}) cannot be passed all the time with the use of 
entanglement by sending just one classical bit.

\section{Remote State Measurement of photon}

In an interesting protocol called ``classical teleportation'' of a qubit 
it is aimed to simulate any possible measurement on a qubit known to Alice but
unknown to Bob using hidden variables and classical communications \cite{cgm}.
Remote state measurement (RSM) protocol is an outcome of quantum version of 
the above protocol. At a first glance, it appears that in our RSP scheme 
one can remotely prepare an arbitrary 
{\em known} photon state one half of the time, so Bob might not be able to 
simulate the measurement statistics all the time as Bob cannot get a {\em 
unknown} photon from the orthogonal photon state. However, a little thought
shows that there is no problem with Bob for simulating the measurement 
statistics on the complement photon. This 
is because the quantum mechanical probabilities and transition probabilities 
are invariant under unitary and anti-unitary operations which is the famous 
Wigner's theorem. This says that for any two non-orthogonal states
if $\ver \la \Psi \ver \Phi \ra \ver^2 = \ver \la \Psi' \ver \Phi' \ra 
\ver^2$ then $\ver \Psi' \ra, \ver \Phi' \ra$ are related to $\ver \Psi \ra, 
\ver \Phi \ra$ either by unitary or anti-unitary transformations. For 
example, if Bob wants to measure an observable $({\bf b}.\sigma)$,
with the projection operator $P_{\pm}({\bf b}) = \frac{1}{2}(1 \pm {\bf b}. 
\sigma )$,
then the probability of measurement outcome in the state $\rho_{\Psi} 
= \ver \Psi \ra \la \Psi \ver = \frac{1}{2} (I + {\bf n}. \sigma) $ is given by
\begin{eqnarray}
P_{\pm}( \rho_{\Psi} ) = {\rm tr}(P_{\pm}({\bf b}) \rho_{\Psi} ) =
\frac{1}{2}(1 \pm {\bf b}. {\bf n} ).
\end{eqnarray}
But suppose Bob gets
$\rho_{\Psi_{\perp}} = \ver \Psi_{\perp} \ra \la \Psi_{\perp} \ver =
\frac{1}{2} (I - {\bf n}. \sigma) $. In this case
the measurement gives a result
\begin{eqnarray}
P_{\pm}( \rho_{\Psi_{\perp}} ) = {\rm tr}(P_{\pm}({\bf b}) 
\rho_{\Psi_{\perp}}) = \frac{1}{2}(1 \mp {\bf b}. {\bf n} ).
\end{eqnarray}
The probabilistic outcomes in (7) and (8) are different.
However, Bob can always chose his apparatus (by reversing the direction 
of ${\bf b}$) such that he can make $P_{\pm}( \rho_{\Psi} )= 
P_{\pm}( \rho_{\Psi_{\perp}} )$.
Note that Bob cannot reverse the direction of ${\bf n}$ but can in principle
reverse the direction of ${\bf b}$. So even if Bob cannot get a qubit 
from a complement qubit (half of the time) still he can get the
same measurement outcomes from it. Therefore, Bob can always simulate 
efficiently the statistics of his measurements on a qubit {\em known} to Alice
but {\em unknown} to him, provided they share an EPR pair and communicate
one classical bit. In the classical case, if Alice and Bob do not share 
entanglement they
need to send $2.19$ bits on the average \cite{cgm}. 

Bob can also simulate probabilistic outcomes via generalized measurements 
such as POVMs $F_{\mu}$ with $\sum_{\mu} F_{\mu}=1$. They can be described 
by the elements $F_{\mu}= \frac{1}{2} (|{\bf f}_{\mu}|I + 
{\bf f}_{\mu}. \sigma) $, where
${\bf f}_{\mu}$'s are vectors on Poincare sphere \cite{cgm}. The probability of
measurement outcome in the state $\rho_{\Psi}$ and $\rho_{\Psi_{\perp}}$ 
are given by
$P_{\mu}(\rho_{\Psi}) = {\rm tr}(F_{\mu} \rho_{\Psi}) = 
\frac{1}{2} (|{\bf f}_{\mu}| + {\bf f}_{\mu}. {\bf n})$ and 
$P_{\mu}(\rho_{\Psi_{\perp}}) = {\rm tr}(F_{\mu} \rho_{\Psi_{\perp}}) = 
\frac{1}{2} (|{\bf f}_{\mu}| - {\bf f}_{\mu}. {\bf n})$,
respectively. These outcome are different, but Bob can flip the vectors
${\bf f}_{\mu}$ to make these simulations identical as if he has a qubit
at his disposal. This is possible again with one unit of entanglement and
one classical bit, whereas classically, if Alice and Bob share hidden 
variables to simulate POVM outcomes Alice and Bob have to communicate
on the average $6.38$ bits (counting forward and backward communications) 
\cite{cgm}. Thus use of quantum entanglement saves $5.38$ bits and 
backward communication in remote simulation of POVMs.

\section{Remote Joint state measurement}

Next we ask the question: Can Alice ask Bob to simulate any joint 
measurement on two unknown photon states with the use of one ebit 
and one classical bit? The general answer is `no'.
Suppose Bob at his disposal has 
another photon in an unknown quantum state $\ver \Phi \ra= 
a\ver H\ra + b\ver V \ra$, with $\rho_{\Phi}= \frac{1}{2}(I+ {\bf m}.\sigma)$,
in addition to a photon that Alice would be
preparing. Now Bob wants to do a joint measurement on the 
state of the first photon (being in $\ver \Psi \ra$ or in 
$\ver \Psi_{\perp} \ra$) and the second photon in the state 
$\ver \Phi \ra$.  Suppose the joint measurement operation is an entangled
operator $\Pi$ given by 
\begin{eqnarray}
\Pi = \frac{1}{4}(I\otimes I+ ({\bf r}.\sigma)\otimes I +
I\otimes ({\bf s}.\sigma)+\sum_{ij} t_{ij} \sigma_i \otimes \sigma_j)
\end{eqnarray}
where ${\bf r}$ and ${\bf s}$ are real vectors and $t_{ij}$ are real 
coefficients. In terms of density matrices, Bob's two photons can be 
in a state (when Alice gets $\ver 1 \ra$)
\begin{eqnarray}
\rho_{\Psi} \otimes \rho_{\Phi}= \frac{1}{4}\big[I\otimes I+ 
({\bf n}.\sigma)\otimes I +
I\otimes ({\bf m}.\sigma)+ ({\bf n}.\sigma) \otimes ({\bf m}.\sigma) \big]
\end{eqnarray}
or in a state (when Alice gets $\ver 0 \ra$) 
\begin{eqnarray}
\rho_{\Psi_{\perp} } \otimes \rho_{\Phi}= \frac{1}{4}\big[I\otimes I -
({\bf n}.\sigma)\otimes I +
I\otimes ({\bf m}.\sigma)- ({\bf n}.\sigma) \otimes ({\bf m}.\sigma) \big]
\end{eqnarray}
The question is can Bob make the following two probabilities equal, i.e.,
if ${\rm tr}\big[\Pi \rho_{\Psi} \otimes \rho_{\Phi}]= 
{\rm tr}\big[\Pi \rho_{\Psi_{\perp}} \otimes \rho_{\Phi}]$?
Explicitly, they are given by
\begin{eqnarray}
{\rm tr}\big[\Pi \rho_{\Psi} \otimes \rho_{\Phi}]&=& 
\frac{1}{4}\big[1+{\bf r}.{\bf n}+ {\bf s}.{\bf m}+\sum_{ij}t_{ij} {\bf n}_i
{\bf m}_j \big] \nonumber\\
{\rm tr}\big[\Pi \rho_{\Psi_{\perp}} \otimes \rho_{\Phi}]&=& 
\frac{1}{4}\big[1-{\bf r}.{\bf n}+ {\bf s}.{\bf m}-\sum_{ij}t_{ij} {\bf n}_i
{\bf m}_j \big]
\end{eqnarray}
In general these two probabilities are not equal and there is no way for
Bob to make them equal either. This is because Bob can only manipulate with his
measuring device described by parameters ${\bf r}$ and ${\bf s}$. 
By flipping the sign of ${\bf r}$ and ${\bf s}$ he cannot make these 
probabilities identical.
However, examination of the above equation reveals 
that if the additional quantum state $\ver \Phi \ra$ is {\em known}
to Bob, then he can flip ${\bf m}$ and ${\bf r}$ to make these probabilities 
equal. Thus remote joint state measurement is possible on an arbitrary photon
that Alice wanted, together with a known photon state.

\section{Further generalisation and open questions}

Remote state preparation and measurement protocols have brought out various
features of quantum and classical resources used in quantum communication.
Unlike in teleportation where resources are fixed for the task, here it
is possible to have trade-offs. In last three years, important progresses have
been made in understanding these trade-offs.
Soon after the exact RSP protocol, Lo has \cite{hklo} conjectured that if 
Alice wants to prepare remotely an arbitrary qubit it may still require 
two classical bits as in the case of quantum teleportation. 
Bennett {\it et al} have generalised RSP for arbitrary qubits, higher 
dimensional Hilbert spaces and also RSP of entangled systems. If one 
does not restrict the number of entangled pairs used, 
then asymptotically one can prepare an arbitrary qubit with one classical 
bit \cite{betal}. Subsequently, Devetak and Berger have proposed a low 
entanglement RSP protocol \cite{db} for arbitrary quantum states. 
The exact and minimal resource consuming RSP protocol
is generalised to higher dimension by  Zeng and Zhang \cite{zeng}.
It was found that it is not possible to have RSP in arbitrary higher 
dimension with minimal resources. There are restrictions on the dimension of
the Hilbert space for which RSP can be realized. 
Leung and Shor have given a stronger proof of Lo's conjecture for RSP of
arbitrary quantum state \cite{debi}. Remote preparation of ensemble of
mixed states has been studied by Berry and Sanders \cite{berry}. 
In addition, as a first step the exact RSP and RSM protocol for qubit 
\cite{akp} have been implemented using NMR devices \cite{peng,peng1} 
over atomic distances. However, 
long distance implementation of RSP of special and arbitrary qubits 
and RSM of arbitrary qubits would be very welcome in future. Even 
though RSP could 
be generalised to higher dimensional Hilbert space, similar generalisation 
of RSM to higher dimensional quantum systems is still elusive. It may be even
impossible! So in that sense only qubits (photons) enjoy the RSM protocol.

\vskip 1cm

{\it Invited talk in 6th International 
Conference PHOTONICS-2002, held from 16th-18th Dec. 2002, in 
TIFR, Mumbai, India.}

\vskip .5cm

\nopagebreak
\renewcommand{\baselinestretch}{1}

\end{document}